\documentclass[aps, 10pt, prb, showkeys, twocolumn, superscriptaddress]{revtex4-2}
\usepackage{graphicx, xcolor, amssymb, amsmath, amsfonts, enumitem}
\usepackage[utf8]{inputenc}
\usepackage[normalem]{ulem}
\usepackage[colorlinks=true, linkcolor=blue, citecolor=blue, urlcolor=blue]{hyperref}
\usepackage{url, soul, ulem, textgreek, bm, comment}
\usepackage[dvipsnames]{xcolor}
\usepackage{appendix}

\newcommand{\apctpadd}{Asia Pacific Center for Theoretical Physics, Pohang, Gyeongbuk, 37673, Republic of Korea}
\newcommand{\postechadd}{Department of Physics, Pohang University of Science and Technology, Pohang, Gyeongbuk 37673, Korea}

\begin{document}

\author{Sanghoon Lee}
\email{sanghoon.lee@apctp.org}
\affiliation{\apctpadd}

\author{Tilen \v{C}ade\v{z}}
\email{tilen.cadez@apctp.org}
\affiliation{\apctpadd}

\author{Kyoung-Min Kim}
\email{kyoungmin.kim@apctp.org}
\affiliation{\apctpadd}
\affiliation{\postechadd}

\date{\today}

\title{Structural constraints on mobility edges in one-dimensional quasiperiodic systems}

\begin{abstract}
Mobility edges commonly arise in one-dimensional quasiperiodic systems once exact self-duality is broken, yet their origin is typically understood only at the level of individual Hamiltonians.
Here we show that mobility edge positions are not independent spectral features of individual Hamiltonians, but are structurally constrained across quasiperiodic Hamiltonians related by an isospectral duality.
Using a bichromatic Aubry--Andr\'e model as a minimal setting, we demonstrate that this constraint is encoded in an exact identity for Lyapunov exponents derived from the Thouless formula.
As a consequence, the mobility edge positions are restricted to a reduced set of energies.
In the self-dual limit, these mobility edge positions coincide at a single localization--delocalization transition.
This structural constraint enforces a linear critical scaling of the physical Lyapunov spectrum near the self-dual point.
Numerical results confirm a critical exponent consistent with the standard Aubry--Andr\'e value of \(\nu = 1\), while simultaneously revealing a novel, non-universal energy-dependent prefactor.
\end{abstract}

\maketitle

\section{INTRODUCTION}


Quasiperiodic systems (QPS) provide a unique setting for studying novel localization, transport, and topological phenomena that beyond condensed matter physics span multiple scientific fields involving wave systems. Like disordered systems, quasiperiodic systems also break translational invariance, but through deterministic quasiperiodicity rather than randomness. In conventional disordered systems belonging to the orthogonal symmetry class, arbitrarily weak uncorrelated disorder localizes all single-particle states in one and two dimensions~\cite{abrahams1979scaling, mackinnon1981oneparameter}, while a localization-delocalization transition occurs only in three dimensions and higher~\cite{anderson1958absence}. In contrast, the quasiperiodic potential enables the localization-delocalization transition already in one dimension~\cite{aubry1980analyticity}.

The paradigmatic example of a QPS is the Aubry--Andr\'e (AA) model~\cite{aubry1980analyticity}, which describes a particle hopping on a one-dimensional lattice in the presence of a quasiperiodic potential. It exhibits an exact self-duality that relates real and momentum space and leads to a localization-delocalization transition that occurs at the same critical disorder for all eigenstates. Generally, modifications of the AA model lead to the emergence of mobility edges. This phenomenon has been demonstrated in a variety of quasiperiodic lattice models, including systems with longer range hopping terms~\cite{biddle2009localizationA, biddle2010predicted, biddle2011localizationB,  saha2019anomalous, deng2019power-law}, modified potentials~\cite{dassarma1988mobility, meir1988localization, varga1992power-law, ganeshan2015nearest, zhang2022lyapunov, goncalves2023rg} such as multiple incommensurate harmonic potentials~\cite{soukoulis1982localization, hiramoto1989scaling, ribeiro2013multichromatic, boers2007mobility, li2017mobility} as well as the combination of both long range hopping and modified potential~\cite{gopalakrishnan2017selfdual, goncalves2023critical}, coupled chains~\cite{rossignolo2018localization, li2020mobility}, in the presence of additional nonlinear modulations~\cite{utesov2025mobility} as well as interactions~\cite{
Iyer2013mbl, Mondaini2015mbl, znidaric2018interaction, 
Xu2019butterfly, Doggen2019mbl, Aramthottil2021fss, Vu2022mbl, goncalves2024moire} or in flatband systems~\cite{bodyfelt2014flatbands, danieli2015FB, lee2023criticalA, lee2023criticalB}. Intense interest in QPS is due to recent significant experimental progress in various platforms: cold atoms in optical lattices~\cite{Roati2008AL, deissler2010delocalization, schreiber2015observation, Bordia2017PeriodicallyDrivingMBL, An2018zigzag,  An2021InteractionsMobilityEdges, luschen2018single, Shimasaki2024KickedQuasicrystal}, photonic lattices~\cite{Lahini2009_Quasiperiodic_Photonic, Kraus2012TopologyQuasicrystals, verbin2013observation, Wang2020Localization}, acoustic systems~\cite{Ni2019Observation, Apigo2019QPAcousticWaveguide} and cavity-polariton devices~\cite{Goblot2020Emergence}.

\begin{figure}[t]
    \centering
    \includegraphics[width=\linewidth]{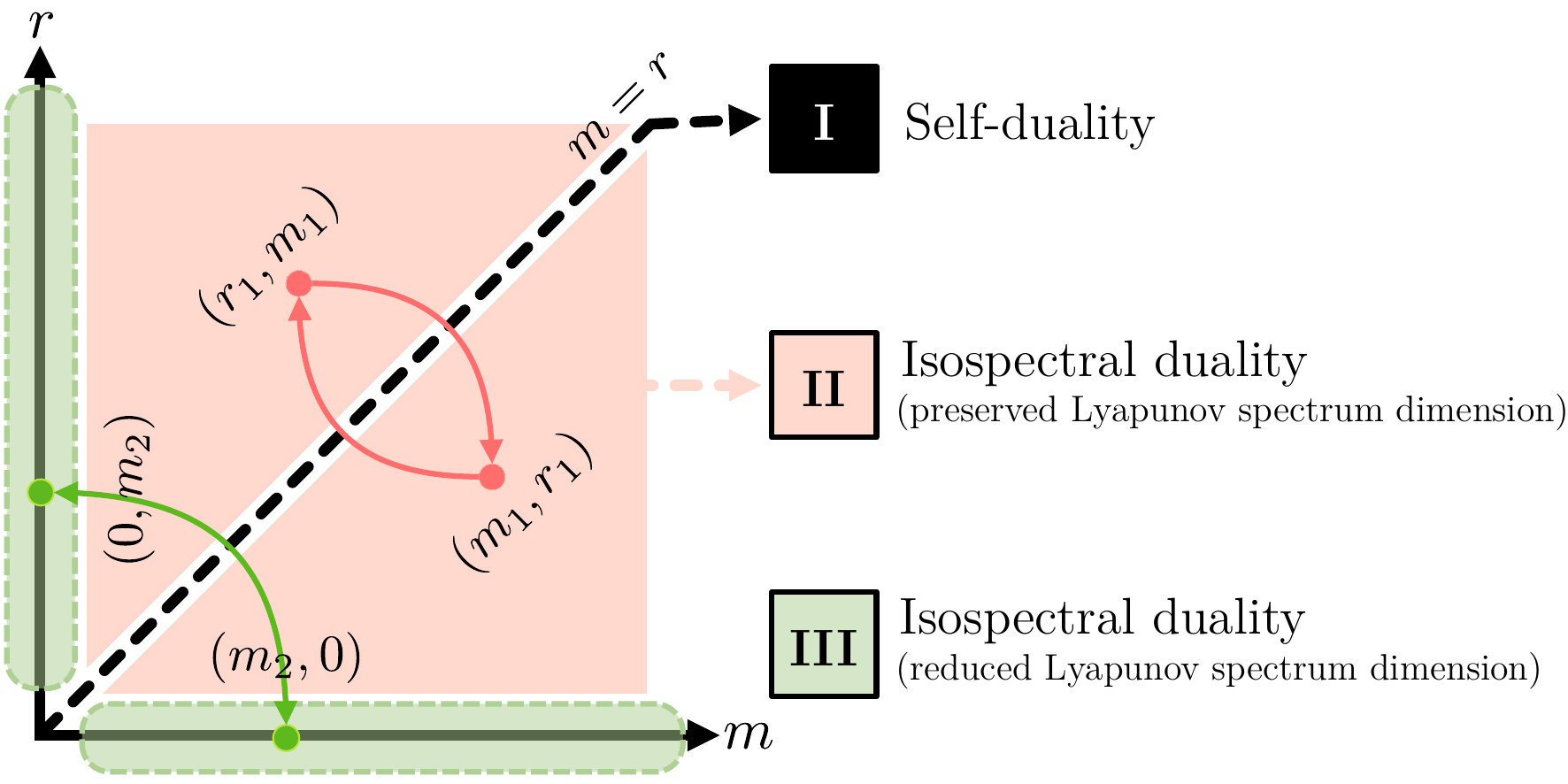}
    \caption{\!\!\!
        Classification of the parameter space in terms of isospectral duality and Lyapunov spectrum structure.
        Region I corresponds to the self-dual manifold.
        Region II represents an isospectral duality with preserved Lyapnuov spectrum dimension, where the number of positive Lyapunov exponents is identical for the dual Hamiltonians.
        Region III denotes the special isospectral regime, in which the Lyapunov spectrum dimension differs between the original and dual systems.
    }
    \label{fig:schema}
\end{figure}

In this work, we adopt a structural view of mobility edges in quasiperiodic systems.
Instead of asking how mobility edges are identified in a single Hamiltonian—corresponding to a single point in the parameter space \((m,r)\)—we ask how they behave across isospectral Hamiltonians that are structurally related [Fig.~\ref{fig:schema}].
We show that when two Hamiltonians are connected by an isospectral duality, their mobility edges are not an independent quantity, but are constrained by an exact relation imposed by the Lyapunov spectrum.
This reveals that mobility edges cannot, in general, be specified independently across such models, but instead obey a structural constraint that becomes visible only when families of dual Hamiltonians are considered.
Furthermore, we find that isospectral dual models can exhibit an energy-dependent scaling behavior that is absent in the standard AA model [at \((m,r)=(0,0)\)].

As a minimal concrete example, we focus on the bichromatic Aubry--Andr\'e (BAA) model, where this structural duality becomes explicit. This structural relation goes beyond the conventional AA self-duality by relating physically distinct Hamiltonians, rather than mapping a model onto itself.
A direct consequence of this relation is that the Lyapunov spectra of the dual Hamiltonians are quantitatively constrained and cannot vary independently, implying that the positions of mobility edges are restricted.
In the self-dual limit, this constraint forces the mobility edge positions to coincide at a single localization--delocalization transition, recovering the AA transition as a special fixed point of the structure.
This enforced coincidence further allows us to analyze the critical scaling at the transition and to show that the observed critical behavior follows from the underlying structural constraint.

The manuscript is organized as follows. In Sec.~\ref{sec:model} we introduce the BAA model and demonstrate the constraints on mobility edges in Sec.~\ref{sec:mobility_edge_constaint}.
Critical behavior at self-dual line of the model is explored numerically in Sec.~\ref{sec:critical-exponent}.
Finally, a discussion is presented in Sec.~\ref{sec:conclusions}.

\section{MODEL}
\label{sec:model}

We consider a one-dimensional quasiperiodic tight-binding model of the BAA type, which serves as the starting point of our analysis.
The Hamiltonian \(H\) takes the form
\begin{align}\label{eq:H_model}
    (H\psi^{(H)})_{n}
    &= g\left[ 2\cos(2\pi\alpha n) + 2m\cos(4\pi\alpha n) \right] \psi^{(H)}_{n} \notag\\[3pt]
    &+ \left( \psi^{(H)}_{n+1}+\psi^{(H)}_{n-1} + r(\psi^{(H)}_{n+2}+\psi^{(H)}_{n-2}) \right),
\end{align}
where \(g\) controls the relative strength between the onsite potential and the hopping terms.
Throughout this work, we fix the irrational frequency to \(\alpha = (\sqrt{5}-1)/2\).
The parameters \(m\) and \(r\), respectively, characterize the relative strength of the second harmonic in the onsite potential and the next-nearest-neighbor hopping, while the nearest-neighbor hopping amplitude is set to unity \(t_{1} = 1\), fixing the overall energy scale.

A discrete Fourier transform maps Eq.~\eqref{eq:H_model} to a dual lattice representation in which the roles of the hopping amplitudes and the onsite potential amplitudes are exchanged.
This mapping preserves both quasiperiodic structure and the energy spectrum of the model.
Therefore, we refer to the resulting representation as an \textit{isospectral dual representation}.

In the following, we consider two Hamiltonians related by this isospectral duality.
The original Hamiltonian is given by \(H\) in Eq.~\eqref{eq:H_model}, while the isospectral dual Hamiltonian \(K\) takes the form
\begin{align}\label{eq:K_model}
    (K\psi^{(K)})_{n}
    &= \left[ 2\cos(2\pi\alpha n) + 2r\cos(4\pi\alpha n) \right] \psi^{(K)}_{n} \notag\\[3pt]
    &+ g\left( \psi^{(K)}_{n+1}+\psi^{(K)}_{n-1} + m(\psi^{(K)}_{n+2}+\psi^{(K)}_{n-2}) \right).
\end{align}
Under this isospectral dual construction, the Hamiltonian \(H\) and \(K\) share the same energy spectrum, but differ in the assignment of kinetic and potential terms.

Furthermore, an AA-like self-dual point arises when the isospectral duality reduces to a self-mapping of the model.
This condition is realized for \(m = r\), where the two Hamiltonians \(H\) and \(K\) coincide at \(g = 1\).

\section{CONSTRAINT ON MOBILITY EDGES}
\label{sec:mobility_edge_constaint}

In this section, we show that mobility edge positions of quasiperiodic Hamiltonians related by an isospectral duality cannot vary independently.
Instead, the allowed locations of mobility edges are constrained across the dual pair, reflecting a structural relation between their Lyapunov spectra.
This constraint is formulated at a structural level and does not rely on model-specific details.
Instead, the argument rests only on three ingredients: the existence of an isospectral structural duality relating two quasiperiodic Hamiltonians, the applicability of a Thouless formula for the sum of positive Lyapunov exponents of the associated transfer matrices, and the definition of mobility edges as the energy at which the smallest positive Lyapunov exponent vanishes.

Localization properties of one-dimensional quasiperiodic systems can be described using Lyapunov exponents of the corresponding transfer matrix, as ensured by the Oseledets' theorem~\cite{oseledets1968multiplicative}.
For the present model, the size of the transfer matrix is \(4\times 4\) and admits two positive Lypaunov exponents, ordered as \(\gamma_{1} \geq \gamma_{2} \geq 0\).
The physical localization length is determined by the smallest positive Lyapunov exponent \(\xi = 1/\gamma_{2}\)~\cite{kramer1993localization}, and a mobility edges \(E_{c}\) are defined by the condition at which physical Lyapunov exponent vanishes, \(\gamma_{2}(E_{c}) = 0\).
For notational simplicity, we collect the parameters into the vector form \(\boldsymbol{\lambda} = (g,m,r)\).
The isospectral duality introduced in Sec.~\ref{sec:model} can be represented by the following dual map \(\mathcal{D} : \boldsymbol{\lambda} \to \mathcal{D}(\boldsymbol{\lambda})\) defined by \(\mathcal{D}(\boldsymbol{\lambda}) = (1/g, r, m)\).

For quasiperiodic systems with higher-rank transfer matrix, the sum of positive Lyapunov exponents satisfies a Thouless formula~\cite{thouless1972relation, chapman2015localization, oliveira2022kotani}.
For the Hamiltonian \(H\), it takes the form
\begin{align}\label{eq:thouless}
    \Gamma_{\!\!H}(E;\boldsymbol{\lambda})
    &= \gamma^{H}_{1}(E;\boldsymbol{\lambda}) + \gamma^{H}_{2}(E;\boldsymbol{\lambda}) \notag\\
    &= \int_{\mathbb{R}}\ln|E - x|dN^{H}_{\boldsymbol{\lambda}}(x) - \ln|r|,
\end{align}
where \(N_{\boldsymbol{\lambda}}^{H}\) is the integrated density of states (IDOS), defined as the thermodynamic limit of the normalized eigenvalue counting function.

\begin{figure}[t]
    \centering
    \begin{minipage}[t]{0.49\linewidth}
        \centering
        \includegraphics[width=\linewidth]{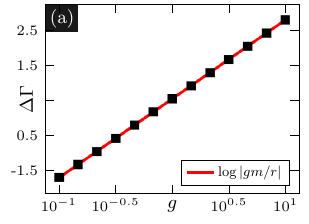}
    \end{minipage}
    \hfill
    \begin{minipage}[t]{0.49\linewidth}
        \centering
        \includegraphics[width=\linewidth]{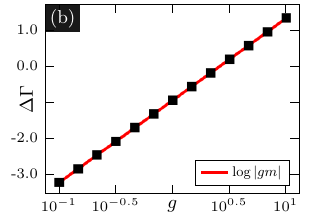}
    \end{minipage}
    \caption{\!\!\!
        Difference of Lyapunov-exponent sums \(\Delta\Gamma\) as a function of the parameter \(g\).
        (a) For \(E = 1.234\) with \(m = 1.2\) and \(r = 0.7\), the numerical data (black) for \(\Delta \Gamma\) coincide exactly with \(\ln|gm/r|\) (red), as given in Eq.~\eqref{eq:diff_Gamma}.
        (b) For \(E = 1.234\) and \(m = 0.4\) at \(r = 0\), the numerical results (black) agree with \(\ln|gm|\) (red) in accordance with Eq.~\eqref{eq:diff_Gamma_special}.
        The exact agreement in both panels demonstrates that \(\Delta\Gamma\) is independent of energy and is governed solely by \(g\), as predicted by the theory.
    }
    \label{fig:isospectral-duality}
\end{figure}

Isospectral duality implies a scaling relation between the spectra of \(H\) and its dual \(K\), leading to the identity
\begin{gather}
    \mathrm{Spec}(H(E;\boldsymbol{\lambda})) = g\cdot\mathrm{Spec}\!\left( K(E/g; \mathcal{D}(\boldsymbol{\lambda})\right)).
\end{gather}
This scaling relation implies that the IDOS follows:
\begin{gather}
    N^{H}_{\boldsymbol{\lambda}}(E) = N^{K}_{\mathcal{D}(\boldsymbol{\lambda})}(E/g).
\end{gather}
Substituting this into the Thouless formula in Eq.~\eqref{eq:thouless} yields the difference \(\Delta\Gamma\) of Lyapunov exponent sums between the \(H\) and its structural dual \(K\),
\begin{gather}\label{eq:diff_Gamma}
    \Delta\Gamma := \Gamma_{\!\!H}(E;\boldsymbol{\lambda}) - \Gamma_{\!\!K}(E/g ; \mathcal{D}(\boldsymbol{\lambda})) = \ln\left|\frac{gm}{r}\right|.
\end{gather}
which is independent of Energy.
Eq.~\eqref{eq:diff_Gamma} constitutes the central result of this section, establishing an exact relation between the Lyapunov exponent sums of the dual Hamiltonians .

To verify this relation numerically, we compute the Lyapunov exponents using a standard transfer-matrix approach.
Fig.~\ref{fig:isospectral-duality} illustrates the resulting energy-independent behavior of \(\Delta\Gamma\).
Throughout this work, the Lyapunov exponents are computed using \(N_{\rm iter} = 150,000\) transfer matrix iterations, discarding the first \(50,000\) steps as burn-in.
A QR decomposition is applied every five steps, and results are averaged over ten realizations.

To make the duality constraint explicit at the localization--delocalization transition, we introduce the difference between the physical Lyapunov exponents of \(H\) and its isospectral dual \(K\),
\begin{gather}\label{eq:diff_phys_le}
    F(E;\boldsymbol{\lambda}) := \gamma_{2}^{H}(E;\boldsymbol{\lambda}) - \gamma_{2}^{K}(E/g;\mathcal{D}(\boldsymbol{\lambda})).
\end{gather}
This quantity will be used to characterize the locations of mobility edges under the isospectral duality.
At mobility edges \(E_c\), the physical Lyapunov exponent vanishes, \(\gamma^{H}_{2}(E_{c}) = 0\).
By isospectral duality, this transition corresponds to mobility edges of the dual Hamiltonian \(K\) at the rescaled energy \(E_{c}/g_{c}\) where \(g_{c} = g(E_{c})\), hence \(\gamma_{2}^{K}(E_{c}/g_{c}) = 0\).
Consequently, the condition \(F(E_{c};\boldsymbol{\lambda}_{c}) = 0\) holds at the transition.
Evaluating Eq.~\eqref{eq:diff_phys_le} at this dual pair of transition points yields the implicit condition
\begin{gather}\label{eq:me_condition}
    \ln\left|\frac{g_{c}m}{r}\right| + \gamma^{K}_{1}(E_{c}/g_{c} \,; \mathcal{D}(\boldsymbol{\lambda}_{c})) - \gamma^{H}_{1}(E_{c};\boldsymbol{\lambda}_{c}) = 0,
\end{gather}
where \(\boldsymbol{\lambda}_{c} = (g_{c},m,r)\).
Eq.~\eqref{eq:me_condition} determines the allowed mobility edge positions, demonstrating explicitly that they are constrained by the isospectral duality.
\begin{figure}[t]
    \centering
    \begin{minipage}[t]{0.49\linewidth}
        \centering
        \includegraphics[width=\linewidth]{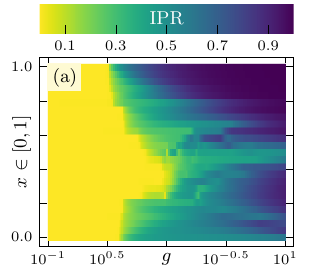}
    \end{minipage}
    \hfill
    \begin{minipage}[t]{0.49\linewidth}
        \centering
        \includegraphics[width=\linewidth]{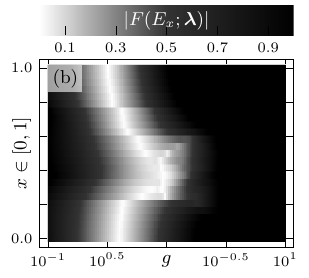}
    \end{minipage}
    \caption{\!\!\!
        Phase diagram of \(H\) at \(r=0\) with \(m=1\), with the parameter sampled over \(g\in [0.1,10]\).
        The vertical axis represents energy parameterized by the IDOS coordinate \(x = N(E)\), discretized into \(N_{\rm bin}=25\) bins, to account for spectral bandwidth variations.
        (a) Inverse participation ratio (IPR) obtained from exact diagonalization on a lattice with \(L = 3200\) sites.
        Low (high) IPR values indicate extended (localized) eigenstates.
        (b) \(|F(E;g)|\) with respect to \(g\), with energies obtained from diagonalization of a finite lattice with \(L = 101\) sites.
        Values with \(|F(E;g)| > 1\) are shown black.
        The mobility edges (white region) follow Eq.~\eqref{eq:me_condition_at_r_0}, demonstrating that they are fully determined by a Biddle--Das Sarma type \(K\).
    }
    \label{fig:SE_model_me}
\end{figure}

As a consistency check, we consider the special case \(r = 0\), where \(H\) reduces to a purely nearest-neighbor quasiperiodic system of Soukoulis--Economou~\cite{soukoulis1982localization} type, characterized by a \(2\times 2\) transfer matrix and a single positive Lyapunov exponent.
In this limit, the Lyapunov-exponent sum coincides with the physical one~\cite{thouless1972relation},
\begin{gather}
    \Gamma_{\!\!H}(E;\boldsymbol{\lambda})
    =
    \int_{\mathbb{R}}\log|E - x|dN^{H}_{\boldsymbol{\lambda}}(x)
    =
    \gamma^{H}_{2}(E;\boldsymbol{\lambda}).
\end{gather}
We emphasize that this limit cannot be obtained by naively substituting \(r = 0\) into Eq.~\eqref{eq:diff_Gamma}.
The logarithmic term \(\ln|r|\) appearing in Eq.~\eqref{eq:diff_Gamma} originates from the finite maximal hopping range associated with the next-nearest-neighbor amplitude \(r\).
When \(r = 0\), the hopping is restricted to nearest neighbors only, and the corresponding contribution is \(\ln|t_{1}|\).
Since we have fixed \(t_{1} = 1\) through out this work, this logarithmic term vanishes.
Eq.~\eqref{eq:diff_Gamma}, then, reduces to
\begin{gather}\label{eq:diff_Gamma_special}
    \gamma^{H}_{2}\!(E;\boldsymbol{\lambda}) - \Gamma_{\!\!K}(E/g;\mathcal{D}(\boldsymbol{\lambda})) = \ln|gm|.
\end{gather}
At mobility edges, this yields the condition
\begin{gather}\label{eq:me_condition_at_r_0}
    F(E_{c};\boldsymbol{\lambda}_{c}) = \ln|g_{c}m| + \gamma^{K}_{1}(E_{c}/g_{c} \,; \mathcal{D}(\boldsymbol{\lambda}_{c})) = 0,
\end{gather}
which explicitly determines the mobility edges of \(H\) through Lyapunov spectrum of its dual Hamiltonian \(K\).

Fig.~\ref{fig:SE_model_me} illustrates this relation for the case \(m = 1\) and \(r = 0\).
The mobility edges, indicated by the white regions, follow Eq.~\eqref{eq:me_condition_at_r_0} and are therefore entirely determined by the properties of the dual Hamiltonian \(K\).
In this limit, \(K\) corresponds to a Biddle--Das Sarma~\cite{biddle2011localizationB} type model.

\section{CRITICAL BEHAVIOR AT SELF-DUAL LINE}
\label{sec:critical-exponent}

We now turn to the critical behavior at the self-dual line \(m=r\), where the Fourier transform maps the Hamiltonian in Eq. \eqref{eq:H_model} onto itself. In this regime, the isospectral duality reduces to a self-duality with a localization--delocalization transition at \(g_{c} = 1\).
We find that the self-dual structure enforces a linear critical scaling of the physical Lyapunov exponent
\begin{gather}
    \gamma_2 \sim |g-1|^{\nu}, \qquad \nu \simeq 1,
\end{gather}
independent on the choice of \(m\).

The origin of this linear scaling can be traced back to the self-dual identity satisfied by the Lyapunov-exponent sum.
On the self-dual manifold, Eq.~\eqref{eq:diff_Gamma} reduces to 
\begin{gather}\label{eq:diff_Gamma_sd}
    \Gamma(E;g)-\Gamma(E/g;1/g)=\ln|g|.
\end{gather}
Near the self-dual point \(g = 1\), the expansion \(\ln|g| = \pm(g-1) + \mathcal{O}((g-1)^{2})\) holds.
Thus, Eq.~\eqref{eq:diff_Gamma_sd} enforces an \(|g-1|\) variation of the Lyapunov-exponent sum.
Along the self-dual manifold, the two dual description belong to the same self-dual family and differ only by a reparametrization of \(g\).
Throughout this section, we therefore use \(g\) as the sole control parameter, omitting the superscripts distinguishing \(H\) and \(K\) for notational simplicity.

To understand how this enforced linear scaling in \(|g-1|\) can arise, we consider the leading critical behavior of the individual positive Lyapunov exponents as \(g \to 1\).
Close to the localization transition, each positive exponent may obey a power-law scaling of the form
\begin{gather}
    \gamma_i(E,g)\sim A_i(E)\,|g-1|^{\nu_i},\quad i\in\{1,2\},
\end{gather}
where \(A_i(E)\) are non-universal, energy-dependent prefactors.
Since the total Lyapunov exponent is given by \(\Gamma = \gamma_{1} + \gamma_{2}\), the linear scaling in \(|g-1|\) enforced by Eq.~\eqref{eq:diff_Gamma_sd} cannot be realized if both positive Lyapunov exponents vanish faster than linearly.
Indeed, if \(\nu_{1,2} > 1\), their sum would also vanish faster than \(|g-1|\), in contradiction with Eq.~\eqref{eq:diff_Gamma_sd}.
Therefore, the self-duality imposes the bound \(\min(\nu_{1},\nu_{2}) \leq 1\).
This bound originates from the coexistence of multiple positive Lyapunov exponents under exact self-duality, in contrast to the single-exponent of the AA model.

\begin{figure}[t]
    \centering
    \begin{minipage}[t]{0.49\linewidth}
        \vspace{0pt}
        \centering
        \includegraphics[width=\linewidth]{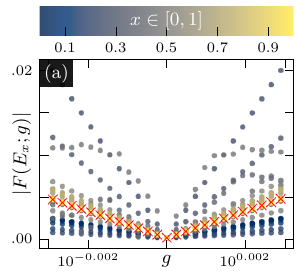}
    \end{minipage}
    \hfill
    \begin{minipage}[t]{0.485\linewidth}
        \vspace{0pt}
        \centering
        \includegraphics[width=\linewidth]{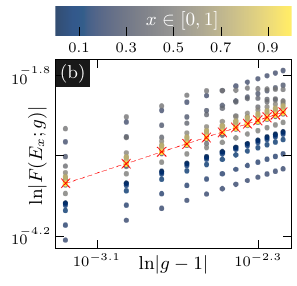}
    \end{minipage}
    \vspace{3mm}
    %
    \begin{minipage}[t]{\columnwidth}
        \vspace{0pt}
        \centering
        \includegraphics[width=\columnwidth]{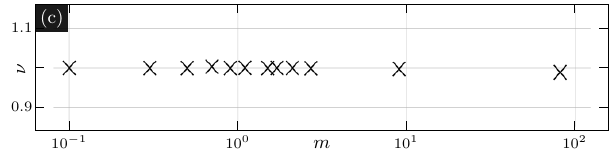}
    \end{minipage}
    \caption{\!\!\!
        Critical behavior of \(F(E;g)\) near the self-dual point.
        The energy is parametrized by the IDOS \(x = N(E)\), discretized into \(N_{\rm bin}=25\) bins such that \(|F(E;g)|\) is plotted as a function of \(g\) for fixed values of \(x\).
        Energies are obtained from diagonalizing a finite lattice with \(L = 151\) sites.
        (a) \(|F(E;g)|\) as a function of \(g\) near \(g = 1\) at \(m=r =0.7\).
        Red crosses indicate the median over \(x\), respectively.
        (b) Log--log plot of \(\ln|F(E;g)|\) versus \(\ln|g-1|\), from which a power-law scaling Eq.~\eqref{eq:power_law} is extracted.
        Data for \(g < 1\) are omitted owing to the symmetry.
        Dashed lines indicate linear fits for the median (red) values.
        (c) Critical exponent \(\nu\) as a function of \(m\), evaluated for \(m=0.1,\,0.3,\,0.5,\,0.7,\,0.9,\,1.1,\,1.5,\,1.7,\,2.1,\,2.7,\,9.0,\) and \(81.0\).
    }
    \label{fig:critical_scaling}
\end{figure}

To determine whether the physical Lyapunov exponent saturates this bound, we numerically analyze the scaling
\begin{gather}
    |F(E;g)| = |\gamma_{2}(E;g)-\gamma_{2}(E/g;1/g)|.
\end{gather}
For \(g > 1\), \(\gamma_{2}(E/g;1/g)\) vanishes, so that \(|F(E;g)|\) probes the critical behavior of \(\gamma_{2}(E;g)\).
Furthermore, self-duality requires that \(|F(E;g)| \to 0\) as \(g\to 1\) for all energies in the spectrum.
This behavior is confirmed in Fig.~\ref{fig:critical_scaling}~(a) at \(m = r = 0.7\), where \(|F(E;g)|\) vanishes across the spectrum as the self-dual point is approached.
We sample energies with respect to the integrated density of states.
To avoid sensitivity to rare-energy outliers in the energy dependence of \(|F(E;g)|\), we characterize the critical scaling using the median over energies, which provides a measure of the typical behavior.
We find a clear power-law dependence as shown in Fig.~\ref{fig:critical_scaling}~(b),
\begin{gather}\label{eq:power_law}
    \mathrm{med}_{E}\{|F(E;g)|\} \sim |g-1|^{\nu},
\end{gather}
exhibiting a clear linear behavior on a log--log scale, yielding \(\nu\simeq 1\) at \(m = r = 0.7\).
Repeating the analysis for different values of \(m=r\), we consistently obtain \(\nu\simeq 1\), as summarized in Fig.~\ref{fig:critical_scaling}~(c).
This demonstrates that the linear critical scaling is a robust consequence of self-duality, rather than a particular choice of \(m\).

\begin{figure}[t]
    \centering
    \begin{minipage}[t]{0.49\linewidth}
        \vspace{0pt}
        \centering
        \includegraphics[width=\linewidth]{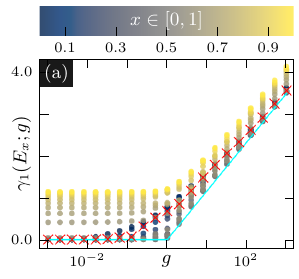}
    \end{minipage}
    \hfill
    \begin{minipage}[t]{0.49\linewidth}
        \vspace{0pt}
        \centering
        \includegraphics[width=\linewidth]{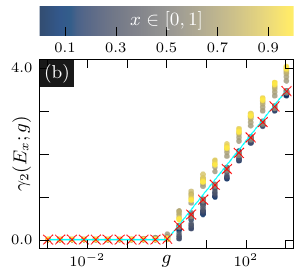}
    \end{minipage}

    \caption{\!\!\!
        Positive Lyapunov exponents \(\gamma_{1}\) and \(\gamma_{2}\) of \(H\) as a function of \(g\) for \(m =r=0.7\).
        The energy is parametrized by the IDOS \(x = N(E)\) discretized into \(N_{\rm bin}=25\) bins, such that \(\gamma_1(E;g)\) is plotted as a function of \(g\) for fixed values of \(x\).
        Energies are obtained from diagonalizing a finite lattice with \(L = 100\) sites.
        (a) \(\gamma_{1}(E;g)\) and (b) \(\gamma_{2}(E;g)\) shown over a wide range of \(g\).
        The red crosses indicate the median, while solid line indicates logarithmic growth \((\ln|g|)/2\).
    }
    \label{fig:large_g_asymptotic}
\end{figure}

Having established the critical scaling exponent \(\nu\simeq 1\) near self-duality, we examine the Lyapunov spectrum deep in the localized regime \(g \gg 1\).
In the AA model, the physical Lyapunov exponent is known exactly as \(\gamma(E;g) = \ln|g|\) for all \(g > 1\), independent of energy.
Motivated by this exact result, we compute the Lyapunov exponents of the BAA model over a wide range of large \(g\).
As shown in Fig.~\ref{fig:large_g_asymptotic}, the smallest positive Lyapunov exponent \(\gamma_{2}(E;g)\) exhibits a logarithmic growth proportional to \((\ln|g|)/2\) throughout the localized regime.
This behavior is characterized by its median over energies.
The larger exponent \(\gamma_{1}(E;g)\) shows a distinct behavior at intermediate \(g\) but converges to the same logarithmic scaling \(\gamma_1\sim(\ln|g|)/2\) for \(g \gg 1\).
This convergence reflects the multi-exponent structure of the BAA transfer matrix, in contrast to the single-exponent structure of the AA model.
The emergence of a logarithmic scale at large \(g\) is consistent with Eq.~\eqref{eq:diff_Gamma_sd} and compatible with the linear critical scaling observed near self-duality.

\section{Non-universal energy-dependent scaling functions}
\label{app:energy_resolved_scaling}

In this section, we demonstrate that the relationship of physical Lyapunov exponents between isospectral dual models, $F(E(x);g)$ as defined in Eq. \eqref{eq:diff_phys_le}, is a non-universal, energy-dependent function.
We perform an energy-resolved scaling analysis by parameterizing the energy through IDOS \(x = N(E)\).
For each energy window labeled by \(x\), we fit
\begin{gather}
    |F(E(x);g)| = A(E(x))\,|g-1|^{\nu}, \quad g>1,
\end{gather}
where \(\nu\) is the critical exponent and \(A(E)\) is a non-universal prefactor.

\begin{figure}[t]
    \centering
    \begin{minipage}[t]{0.49\linewidth}
        \centering
        \includegraphics[width=\linewidth]{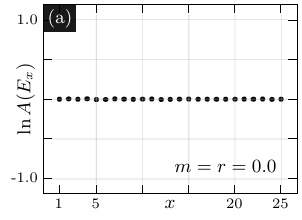}
    \end{minipage}
    \hfill
    \begin{minipage}[t]{0.49\linewidth}
        \centering
        \includegraphics[width=\linewidth]{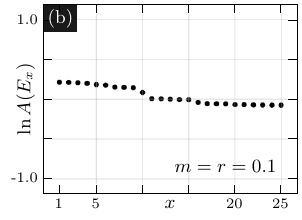}
    \end{minipage}
    \begin{minipage}[t]{0.49\linewidth}
        \centering
        \includegraphics[width=\linewidth]{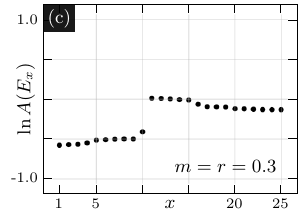}
    \end{minipage}
    \hfill
    \begin{minipage}[t]{0.49\linewidth}
        \centering
        \includegraphics[width=\linewidth]{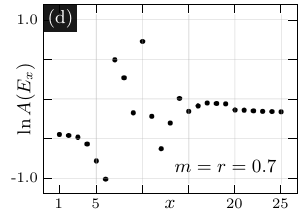}
    \end{minipage}

    \caption{\!\!\!
        Energy-resolved scaling analysis.
        For each energy window parameterized by IDOS \(x=N(E)\) discretized into \(N_{\rm bin}=25\) bins, we extract the prefactor \(\ln A(E)\).
        Panels (a)--(d) correspond to different values of \(m = r\).
        While \(\nu_{x}\) remains nearly constant and close to \(\nu\simeq 1\), the prefactor exhibits energy dependence once the system deviate from the AA model in (a).
    }
    \label{fig:app_energy_resolved_scaling}
\end{figure}

Fig.~\ref{fig:app_energy_resolved_scaling} shows the resulting prefactor \(\ln A(E)\) as function of the energy coordinate \(x\), for several values of \(m = r\).
Especially, the system corresponds to the AA model when \(m = r = 0\).
While \(\nu_{x}\) remains essentially constant over the entire energy range and is always close to \(\nu\simeq 1\), the prefactor \(A(E)\) exhibits an energy dependence once the systems deviate from the AA model, reflecting non-universal features as shown in Fig.~\ref{fig:app_energy_resolved_scaling}.
This separation between a robust exponent and an energy-dependent prefactor demonstrates that the observed critical behavior is not an artifact of a particular energy window or fitting procedure, but instead follows from the underlying structural constraint of the model.
Furthermore, this behavior provides a clear distinction from the AA model, in which such an energy-resolved structure is absent.

\begin{figure}[t]
    \centering
    \begin{minipage}[t]{0.49\linewidth}
        \vspace{0pt}
        \centering
        \includegraphics[width=\linewidth]{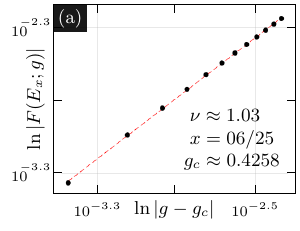}
    \end{minipage}
    \hfill
    \begin{minipage}[t]{0.49\linewidth}
        \vspace{0pt}
        \centering
        \includegraphics[width=\linewidth]{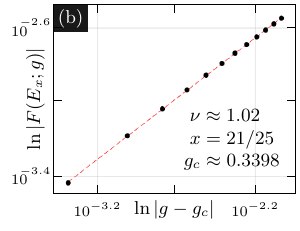}
    \end{minipage}

    \caption{\!\!\!
        Scaling behavior of \(|F(E;g)|\) near the mobility edges for two different energy slices, corresponds to the 6th~(a) and the 21st~(b) energy bins from Fig.~\ref{fig:isospectral-duality}.
        Although the scaling window is reduced in the vicinity of the mobility edges, the observed behavior remains consistent with \(\nu\simeq 1\).
    }
    \label{fig:mobility_edge_critical_exponent}
\end{figure}

To further examine the robustness of the scaling exponent in the presence of mobility edges, we analyze the behavior of \(|F(E;g))\) in the vicinity of the transition for selected energy slices.
Fig.~\ref{fig:mobility_edge_critical_exponent}~(a) and (b) show the results for two different energy windows corresponding to the 6th and the 21st energy bins, respectively, from Fig.~\ref{fig:isospectral-duality}.
In both cases, \(|F(E;g)|\) exhibits a clear power-law scaling on \(|g - g_{c}|\), even though the accessible scaling window becomes narrower near the mobility edges.

Despite this reduced window, the extracted scaling behavior remains consistent with
\begin{gather}
    |F(E;g)| \sim |g-g_c|^{\nu} , \quad\nu \simeq 1.
\end{gather}
This demonstrates that the critical scaling identified in the main text persists locally across the spectrum.
This result further supports the interpretation that the observed critical behavior is governed by the underlying structural constraint.

\section{DISCUSSION}
\label{sec:conclusions}

In this work, we have introduced a structural view on mobility edges in one-dimensional quasiperiodic systems.
Rather than treating mobility edges as isolated spectral features of individual Hamiltonians, we demonstrated that their positions are constrained across Hamiltonians related by an isospectral duality.
Using the BAA as a minimal setting, we derived an exact identity for the sum of Lyapunov exponents based on the Thouless formula.
This identity imposes a nontrivial constraint on the allowed locations of mobility edges, showing that they cannot vary independently between dual Hamiltonians.
In the self-dual limit, we recover the AA critical point as a special case of a more general structural relation.

We further showed that this structural constraint enforces a linear critical scaling of the physical Lyapunov exponent near self-duality, yielding a robust exponent \(\nu\simeq 1\).
Importantly, this behavior does not rely on model-specific parameters but follows directly from the coexistence of multiple positive Lyapunov exponents and the self-dual structure of the system.
Our energy-resolved analysis demonstrates that while the critical exponent remains universal, non-universal prefactors exhibit energy dependence, providing a clear distinction from the AA model.

The model presented here is well-suited for experimental realization using cold atoms in optical lattices~\cite{Roati2008AL, deissler2010delocalization, schreiber2015observation, Bordia2017PeriodicallyDrivingMBL, An2018zigzag, An2021InteractionsMobilityEdges, luschen2018single, Shimasaki2024KickedQuasicrystal}. 
Specifically, the next-nearest-neighbor hopping can be implemented and tuned in engineered synthetic zigzag lattices, where mobility edges may be generated via quasiperiodic potentials or homogeneous flux~\cite{An2018zigzag}. 
The positions of these mobility edges can be determined through the participation ratio, derived from microscopic measurements of site populations for high- and low-energy eigenstates that are accessible via absorption imaging following a time-of-flight period~\cite{An2018zigzag, An2021InteractionsMobilityEdges}. 
Furthermore, the localized phase, the delocalized phase, and the mixed phase with mobility edges can be distinguished by monitoring the density imbalance between even and odd sites alongside the global expansion of the atomic cloud~\cite{luschen2018single}. 

While our analysis was carried out for a bichromatic AA model, the underlying structural arguments do not depend on the specific number of harmonic components. Indeed, the isospectral duality persists in the generalized multichromatic model
\begin{align}
    (H\psi^{(H)})_{n} = & \; g \left[ 2\sum_{w=1}^{N_w} m_w \cos(2w\pi\alpha n) \right] \psi^{(H)}_{n} \notag\\
    & + \sum_{w=1}^{N_w} r_w \left( \psi^{(H)}_{n+w} + \psi^{(H)}_{n-w} \right),
\end{align}
where \(m_1=1\) and \(r_1=1\). This demonstrates that the exact identity for Lyapunov exponents and the resulting constraints on mobility edges apply to a broader class of quasiperiodic tight-binding systems.
In particular, the emergence of contrained mobility edges follows from relations between dual Hamiltonians rather than from model-specific details.
An interesting direction for future work is to investigate how these structural constraints extend beyond the static Hermitian setting considered here, for example to non-Hermitian or driven quasiperiodic systems, where the notion of duality and Lyapunov spectra may exhibit qualitatively different behavior.
More generally, we expect that focusing on structural relations between Hamiltonians may offer a systematic perspective to understand mobility edge phenomena beyond individual model studies.
From a theoretical perspective, our findings also raise an open question concerning the extension of Avila global theory~\cite{avila2015global} of quasiperiodic cocycles.
While the global theory provides a complete characterization for \(2\times 2\) quasiperiodic transfer matrices, the appearance of structural constraints in higher-rank transfer matrix settings (\(4\times 4\) and beyond) suggest the analogous concept may exist beyond \(2\times 2\).

\begin{acknowledgments}
S. L., T.\v{C}., and K.-M.K. were supported by an appointment to the JRG Program at the APCTP through the Science and Technology Promotion Fund and Lottery Fund of the Korean Government. This research was supported by the Korean Local Governments - Gyeongsangbuk-do Province and Pohang City. 

\end{acknowledgments}


\bibliographystyle{apsrev4-2}
\bibliography{ref}

\end{document}